\documentclass{svjour3}
\usepackage{amsmath}
\usepackage{amssymb}
\usepackage{amsfonts}
\usepackage{mathrsfs}

\title{The cosmology of a fundamental scalar}

\author{Jonathan Holland \and George Sparling}

\institute{
University of Pittsburgh, Department of Mathematics, 301 Thackeray Hall, Pittsburgh, PA 15216\\\email{jonathan.e.holland@gmail.com}\\TEL: +1-412-624-8375\\FAX: +1-412-8397
}
\begin{document}

\maketitle




\newpage

\begin{abstract}
We observe that the standard homogeneous cosmologies, those of Minkowski, de Sitter, and anti-de Sitter, which form the matrix for the Robertson--Walker scale factor, live naturally as isolated points inside a larger family of conformally flat metrics obtained by allowing a tensor containing the information of conformal symmetry breaking to be more general.  So the standard cosmological metrics are parametrically unstable in this sense, and therefore unphysical.  When we pass to the stable family of perturbed metrics, we immediately encounter a scalar field, which drives the conformal expansion of the universe and which automatically obeys the non-linear sine-Gordon equation.   The Lagrangian for the sine-Gordon equation is a cosine potential agreeing to the fourth order with the potential used in the approach to the generation of mass in gauge theories.   Accordingly we identify our geometric scalar field---actually of the type of an abelian gauge field---with the recently discovered scalar field.  There are two constants in the theory:  the first, named $m$, is positive and defines a mass scale for the universe; the second, named $\Lambda$, is the cosmological constant.  For the space-time to be everywhere non-singular, equivalently for the (strict) dominant energy condition to hold, these constants must obey the inequality $\Lambda > m^2/4$.
\end{abstract}


\section{Introduction}
Recent experimental evidence  strongly supports the case for the existence of a fundamental scalar field in nature.     Although this scalar has been discovered in the context of high energy experiments,\footnote{See Cho \cite{Cho2012} for a survey of the recent experimental evidence from the LHC for the scalar field predicted by Goldstone \cite{Goldstone1961}, Goldsone, Salam, and Weinberg \cite{Goldstone1962}, Englert and Brout \cite{Brout1964}, Guralnik \cite{Guralnik2009}, and Higgs \cite{Higgs}.} it seems rather likely that the same field plays a crucial role in the early evolution of our universe, where again extremely high energies are encountered (Weinberg \cite{Weinberg2008}).

Using the language of twistor theory developed by Penrose \cite{Penrose1967}, \cite{Penrose1976}, Newman \cite{Newman1976}, and others (see, e.g., Penrose and Rindler \cite{PenroseRindler}), we describe a basic formulation of the standard homogeneous cosmologies, those of Minkowski \cite{Minkowski1908}, de Sitter \cite{deSitter1917}, and anti-de Sitter (Bondi  \cite{Bondi1960}), which comprise the template for the scale factor of Robertson \cite{Robertson1935} and Walker \cite{Walker1937}.  In our formulation these cosmologies are unstable under small perturbations and therefore unphysical.  The stable family of perturbed metrics is a perfect fluid solution with cosmological constant (see Hawking and Ellis \cite{HawkingEllis}) whose Lagrangian exactly agrees with the standard sine-Gordon potential familiar from the theory of solitons as developed by Fermi, Pasta, Ulam, and Tsingou \cite{Fermi1955}, Kruskal \cite{Kruskal1964}, and others.  This class of cosmological perfect fluid solutions has a scale invariance---a timelike conformal Killing vector determined by the fluid---making it an excellent candidate to describe the observed accelerating universe (Perlmutter {\em et al} \cite{Perlmutter1999}, Riess {\em et al} \cite{Riess1998}).

Our approach to the problem of describing the cosmological perfect fluid solutions uses a simple generalization of the idea of conformal symmetry breaking with an infinity twistor, which has recently had spectacular success in the context of the description of quantum gravitational scattering amplitudes due to Witten \cite{Witten2004}, Hodges \cite{Hodges2012}, Cachazo \cite{Cachazo2013}, Cachazo, Mason, and Skinner \cite{CachazoMasonSkinner2012}, and Cachazo {\em et al} \cite{Cachazoetal2012}.  Taken together, these developments reveal a fundamental role for the ideas of twistor theory in physics.

\section{Twistor theory of the Robertson--Walker spacetimes}
The big-bang geometries of Robertson and Walker may be summarized by the following metric, in natural units for which the speed of light is unity:
\[ g_{RW} =    S_{RW}^2  g_\Lambda.\]
Here $g_\Lambda$ is a constant curvature metric, a standard homogeneous metric in four dimensions, with one time variable and three spatial variables, whose Einstein tensor is $3\Lambda g_\Lambda$. So $g_\Lambda$ has for its symmetry group either the de Sitter group ($\Lambda > 0$), the anti-de Sitter group ($\Lambda < 0$), or the Poincar\'{e} group ($\Lambda = 0$), each a real Lie group of dimension ten.   Each reference geometry $g_\Lambda$ is conformally flat and the three possible symmetry groups are each subgroups of the conformal group of spacetime, the fifteen-dimensional Lie group $\mathbb{SO}(2, 4)$.   Given $g_\Lambda$,   the precise metric $g_{RW}$ is determined by the Robertson--Walker scale factor $S_{RW}$ which, via the equations of general relativity of Albert Einstein, depends on the details of the specific theory under consideration and is just a function of one real spacetime variable, a cosmic time variable.

The reference geometries $g_\Lambda$ are beautifully described by the twistor theory for conformally flat spacetime; see Penrose \cite{Penrose1967}, \cite{Penrose1969}, Hughston \cite{Hughston1979}, Huggett \cite{Huggett1994}, and Penrose and Rindler \cite{PenroseRindler}.  In this context, the twistor space $\mathbb{T}$ is a four-dimensional vector space over the complex numbers, equipped with a pseudo-hermitian form, denoted $Z.\overline{Z}$, for the twistor $Z$ in $\mathbb{T}$, of signature $(2, 2)$.   The relevant symmetry group can be taken to be $\mathbb{SU}(2, 2)$, the standard spin group (a double cover) of the conformal group  $\mathbb{SO}(2, 4)$. 

Given $\mathbb{T}$, the associated complex conformally flat spacetime is the Grassmanian of all two-dimensional subspaces of $\mathbb{T}$, and is denoted by $\mathbb{M}_{\mathbb{C}}$.  So $\mathbb{M}_{\mathbb{C}}$ is a complex four-manifold.  Let $\mathbb{K}$ be the five-dimensional projective space of $\Omega^2(\mathbb{T})$,  the exterior product of $\mathbb{T}$ with itself.  Then for $\mathbb{X}$ a point of $\mathbb{M}_{\mathbb{C}}$,  put $X = \Omega^2(\mathbb{X})$, the exterior product of $\mathbb{X}$ with itself.   Then $X$ is a point of $\mathbb{K}$ and as $\mathbb{X}$ varies, $X$ traces out the Klein quadric in $\mathbb{K}$: the four dimensional hypersurface in $\mathbb{K}$ whose equation is $X \wedge X = 0$, with $\wedge$ denoting the exterior product.  Conversely, given a point $X$ of the Klein quadric, the space $\mathbb{X}$ is the set of all twistors $Z$ such that $Z \wedge X = 0$.  Finally, the causal geometry of the spacetime $\mathbb{M}$ is then determined by the condition that distinct points $X$ and $Y$ of the Klein quadric are connected by a null geodesic if and only if $X \wedge Y = 0$, if and only if there is a non-zero twistor $Z$ such that $Z \wedge  X = 0$ and $Z \wedge Y = 0$. 

The twistor space also connects directly with analysis and field equations:  specifically let $\alpha$ be a holomorphic one-form defined on a suitable domain in the projective twistor space, the space of one-dimensional subspaces of $\mathbb{T}$ (for example the domain could be the projective image of the collection of twistors $Z$, such that $Z.\overline{Z} > 0$).  Restrict $\alpha$ to $\mathbb{X}$ (which gives a Riemann sphere in the projective twistor space) and integrate it around a suitable contour on the sphere of $\mathbb{X}$.  This gives a number $\phi(\mathbb{X})$.  As $\mathbb{X}$ varies, we recover a solution of the Paneitz \cite{Paneitz} equation on spacetime, a conformally invariant field equation of the type organized in the work of Graham, Jenne, Mason, and Sparling \cite{GJMS}.  In a conformally flat patch, the equation reduces to the vanishing of the square of the ordinary wave operator when acting on the Paneitz field $\phi$.  Mathematically, we have a natural map from the first sheaf cohomology group with values in one-forms of a suitable domain in twistor space to solutions of the Paneitz equation.   

Similarly, twistor theory describes ordinary massless scalar particles, whose field equation in a conformally flat patch is just given by the vanishing of the wave operator,  through the first sheaf cohomology group of twistor space with values in the sheaf of germs of holomorphic functions of degree minus two on twistor space.  Again the solution $\phi(\mathbb{X})$ is obtained in practice by a suitable contour integral performed on the Riemann sphere of $\mathbb{X}$ (see Penrose and Rindler \cite{PenroseRindler}).

The reduction to the group $\mathbb{SU}(2, 2)$ is determined by an element $\epsilon$ of $\Omega^4(\mathbb{T})$, which may be normalized by the requirement that $\epsilon.\overline{\epsilon} = 24$.  Then the reference metrics $g_\Lambda$ have a uniform description in twistor theory:
\[ g_Y =  2 \frac{\epsilon(dX, dX)}{\epsilon(X,Y)^2}.\]
Here $X$ lies on the Klein quadric, so $X \wedge X = 0$ and $Y$ is a non-zero and constant element of the six-dimensional space $\Omega^2(\mathbb{T})$.   In the complex regime there are only two cases according to whether or not $Y \wedge Y$ vanishes.  The restriction to real metrics is given by restricting $X$ so that $X$ equals the dual with respect to $\epsilon$ of its conjugate $\overline{X}$ and similarly  $Y$ equals the dual with respect to $\epsilon$ of its conjugate $\overline{Y}$.   Then $g_Y$ is real and has constant curvature, and $g_Y=g_{\Lambda}$ where $\Lambda=Y.\overline{Y}$ and all three possibilities for the sign of $\Lambda$ occur.  In turn the symmetry group of $g_\Lambda$ is precisely the subgroup of  the conformal group preserving $Y$.   So $Y$ is the vehicle of conformal symmetry breaking in the style of Sparling \cite{Sparling1981}, \cite{Sparling1986}.  It is called the infinity twistor: the points $X$ for which $X\wedge Y = 0$ constitute conformal infinity, which is null, spacelike or timelike, according as $\Lambda = 0$, $\Lambda > 0$, or $\Lambda < 0$, respectively.

In this formalism, presented in detail in the beautiful works of Penrose and Rindler \cite{PenroseRindler}, the real structure emerges somewhat as an afterthought.  Also there is no use for infinity twistors $Y$ such that $Y$ is not real with respect to the conjugate duality operation.

\section{Instability of the metric}

We now present a slightly different way of describing essentially the same data, where reality is built in \emph{a priori}.  We say that a twistor $Z$ is null if and only if $Z.\overline{Z}$ vanishes.  Then $\mathbb{M}$, a conformal compactification of real Minkowski spacetime and a real four-manifold with topology $\mathbb S^1\times \mathbb S^3$,  is the set of totally null elements $\mathbb{X}$ of $\mathbb{M}_{\mathbb{C}}$:  so if $Z \in \mathbb{X} \in \mathbb{M}\subset \mathbb{M}_{\mathbb{C}}$, then $Z$ is null. This condition is equivalent to the condition that $X(\overline{A})$ be a null twistor for any twistor $A$. 

In this approach, the conformal structure of $\mathbb{M}$ is given by the (manifestly real) conformal metric $g = dX.d\overline{X}$.   Under the replacement $X \rightarrow \lambda X$, where $\lambda $ is non-zero and complex and may depend on $X$, the metric $g$ scales as $g\rightarrow |\lambda|^2 g$, as the terms involving the derivative of $\lambda$ vanish identically.  Consequently, the general metric, $g_f$, representing the conformal structure can be given by the formula:  $f(X) g_f =  dX.d\overline{X}$, where $f(X)$ is a positive function, which has the homogeneity property $f(\lambda X) = |\lambda|^2 f(X)$, for non-zero complex $\lambda$.  


The background metrics $g_Y$ of Robertson--Walker and Penrose--Rindler are now given succinctly by our key formula:  
\[    g_Y = \frac{dX.d\overline{X}}{ |X.\overline{Y}|^2}.\]
So here  $f(X) =  |X.\overline{Y}|^2$, the simplest function with the required homogeneity property.   We then have \emph{exact agreement} with Penrose--Rindler provided that $X$ and the infinity twistor $Y$ obey the usual reality condition, as described above.  In the present formalism, however, there is an irresistible generalization:  we allow the infinity twistor $Y$ to be an arbitrary element of $\Omega^2(\mathbb{T})$, \emph{not constrained} by any reality condition.

To understand the nature of this generalization, first note that the tensor $Y$ has two natural real invariants: the invariant $\Lambda=Y.\overline{Y}$, and twice the modulus of its Pfaffian $m^2=2|\epsilon(Y,Y)|$ with $m\ge 0$.  Introduce the quantity $E_Y$, a pseudo-hermitian endomorphism of twistor space, the $(1, 1)$-tensor obtained by taking a single contraction of the $(2, 2)$-tensor product of $Y$ with its conjugate, $Y\otimes\overline{Y}$. Then $H_Y$, the trace-free part of the operator $E_Y$, generates a one-parameter subgroup of $\mathbb{SU}(2, 2)$, so a one-parameter group of conformal symmetries of (compactified) Minkowski spacetime and also a one parameter group of conformal symmetries of the metric $g_Y$.  The operator $E_Y$ satisfies the quadratic equation
$$u^2 - \frac{1}{2}\Lambda u + \frac{m^4}{256}=0$$
this being its minimal polynomial provided that $E_Y$ is not a multiple of the identity.  The equation has both roots real and distinct provided that $|\Lambda| > \frac{m^2}{4}$, and equal roots precisely when the operator $H_Y$ is nilpotent or zero.  In particular, in the limiting case of $\Lambda=m^2/4$, $Y$ corresponds to the infinity twistor for the de Sitter spacetime.

A preferred one-parameter group of conformal symmetries does not exist for  the Penrose--Rindler metrics, for the simple reason that $g_Y$ is of the Penrose--Rindler type, if and only if $Y$ obeys the reality condition of Penrose and Rindler, if and only if  $E_Y$  is a multiple of the identity, if and only if $H_Y$ vanishes identically, giving the trivial one-parameter subgroup.   In the other special case, where $H_Y$ is non-zero, but with square zero, $H_Y$ gives a (Killing) symmetry for  $g_Y$, not just a conformal symmetry: in the cosmological context, as will be discussed below, this is ruled out experimentally and is also unstable under minuscule changes in the tensor $Y$, just like the Penrose--Rindler metrics.

{We now write out the metric $g_Y$ in suitable coordinates as:\footnote{The spacetime coordinates are related to the skew-symmetric forms $X,Y\in\Omega^2(\mathbb T)$ as follows.  In a standard affine patch of twistor space,
$X=\lambda\begin{pmatrix}
-\frac{1}{2}x.x\epsilon^{AB} & i {x^{A}}_{B'}\\
-i{x^{B}}_{A'} & \epsilon_{A'B'}
\end{pmatrix},$ with $\lambda$ a nonzero complex function of $x$.  Here the (Penrose--Weyl spinor) indices $A,B,A',B'$ are two-dimensional and $\epsilon$ is the standard two-dimensional skew-symmetric Levi-Civita symbol used for raising and lowering indices.  Finally, ${x^{A}}_{B'}=x^{AA'}\epsilon_{A'B'}$ where the four-vector $x^{AA'}$ has the explicit matrix representation $x^{AA'}=\frac{1}{\sqrt{2}}\begin{pmatrix}
x_0-x_1&x_2-ix_3\\
x_2+ix_3&x_0+x_1
\end{pmatrix}.$  Generically $\overline{Y}$ can likewise be written
$\overline{Y}=\mu\begin{pmatrix}
-\epsilon_{AB} & {y_{A}}^{B'}\\
-{y_B}^{A'} & \frac{(z-y\cdot y)}{2}\epsilon^{A'B'}
\end{pmatrix}$ for some complex numbers $\mu\not=0$ and $z$, and complex $4$-vector $y$.  By absorbing the imaginary part of $y$ into a translation in $x$, we can arrange without loss of generality that $y$ is also real.  Also, after at worst a dilation, we can take $|\mu|=1$.}
\[ g_Y = \frac{dx.dx}{\left|(x - iy).(x- iy) + z\right|^2} = \frac{dx.dx}{P^2 + Q^2}, \]
\[ P = \Re(z) + x.x - y.y, \hspace{10pt} Q = \Im(z) - 2x.y, \]
\[ P + iQ = (x - iy).(x- iy) + z.\]
Here $x$ is a real Minkowskian four-vector and $.$ is the usual Minkowskian dot product of signature $(1, 3)$.  So $dx.dx$ is the standard Minkowskian metric.  Also $y$ is real four-vector, $i$ is the square root of minus one and $z$ is a complex number, whose real and imaginary parts are denoted $\Re(z)$ and $\Im(z)$, respectively, such that $16|z|=m^2$.}

It is straightforward to work out the Einstein tensor $G_Y$ for this metric, with the result:
\[ G_Y  = (3\Lambda - \tfrac54 H) g_Y  + \tfrac12 h \otimes h, \hspace{10pt} \Lambda =  4(2y.y - \Re(z)).\]
Here $h$ is a co-vector field and $H = g_Y^{-1}(h, h)$ is the squared ``length'' of the co-vector $h$.  The metric $g_Y$ is everywhere nonsingular on the whole of $\mathbb M$ provided that $P$ and $Q$ never simultaneously vanish, which is equivalent to the condition that $y$ be timelike and $|\Lambda| > m^2/4$.  The stronger inequality $\Lambda > m^2/4$ is equivalent to the requirement that $G_Y$ satisfy the (strict) dominant energy condition.

In the Penrose--Rindler special case---characterized here by the vector $y$ being zero---we have $h =0$, from which it follows that $H = 0$ also, giving $G_Y = 3\Lambda g_Y$, the standard case of constant curvature.   The other special case described above has $h \ne 0$ but $H = 0$, for which $h$ is a non-zero null covariantly constant co-vector.   Then the vector field $g_Y^{-1}(h)$ is a null Killing vector and we have spacetimes of the form of the Rindler wedge and zero acceleration, apparently not in agreement with present observations in cosmology, so ruled out experimentally.

Returning to the general case, it is easy to compute the curl of the one-form $h$, the result being zero.  So $h$ is locally a gradient and is hypersurface orthogonal, so \emph{naturally} foliates the four-space by three-manifolds.  To agree with cosmology, we would probably require that $H > 0$, so that the foliation is by space-like hyper-surfaces, analogous to the foliation of Robertson--Walker.   The conformal Killing field determined by $E_Y$ as discussed above is now just the vector field $g_Y^{-1}(h)$.

Henceforth, we shall assume that $\Lambda>m^2/4$.
\section{The soliton}
The one-form $h$, the carrier of the information of the breaking of conformal invariance, has curl zero.  So we can seek to write it as $h = d\psi$, for some scalar $\psi$ (defined up to an additive constant).  Direct calculation of the Einstein tensor using the explicit form of the metric in coordinates (verified in Maple) gives:
\[ h = d\psi, \hspace{10pt} \psi = 2 \arctan\left(\frac{Q}{P}\right), \hspace{10pt} e^{i\psi} =  \frac{P + iQ}{P - iQ}.  \]
So $\psi$ has the character of an abelian gauge-field with group the multiplicative group of complex numbers of modulus one.  Now we have that $g_Y^{-1}(h)$ is a twist-free conformal Killing vector, so we have the relation:
\[ d\otimes d\psi = d\otimes h = S g_Y.\]
Here $d$ is the Levi-Civita connection and $S$ is a scalar field.  Taking a trace,  we derive the relation $4S = \square \psi$ and it becomes of interest to determine the quantity $S$.   Rather stunningly, we calculate $S$ as:
\[ S = 4\Im\left(z e^{-i\psi}\right). \]
After absorbing a suitable real constant into the potential $\psi$, we may write the equation for $\psi$ as:
\[ \square \psi = - m^2 \sin(\psi).\] 
So the potential $\psi$ obeys the sine-Gordon equation introduced by Bour \cite{Bour1862}.  The Lagrangian for the sine-Gordon equation has a cosine potential and the standard Taylor--Maclaurin series for the cosine of $x$ begins $1-\frac{x^2}{2}+\frac{x^4}{24}$, agreeing to this order with the potential used in the approach to the generation of mass in gauge theories.  Accordingly, we identify $\psi$, the driver of the conformal expansion of the universe and the carrier of the information of the breaking of conformal invariance, with the scalar doublet recently discovered in the Large Hadron Collider (Cho \cite{Cho2012}) in the unitary gauge of Weinberg \cite{Weinberg1973}. The constant $m$ defines a mass scale for the universe.

\section{The action}
The field equations are precisely those derived from the natural Lagrangian, coupling together gravity, the cosmological constant, and the sine-Gordon field:
$$\mathscr L = \frac{1}{16\pi G}[(R + 2\Lambda) + \mathscr L_{\text{matter}}].$$
where $G$ is the gravitational constant and
$$\mathscr L_{\text{matter}}=- \frac{1}{2} g^{-1}(d\psi,d\psi) -\frac{m^2}{\hbar^2}\cos(\psi)$$
with $\hbar$ the reduced Planck constant (which we take to be equal to unity).  The space-time has four-volume
$$\frac{1024\pi^3\Lambda}{(16\Lambda^2-m^4)^{3/2}}$$
and the integral of the action is
$$\frac{8\pi^2}{G\sqrt{16\Lambda^2-m^4}}.$$
So a na\"ive quantization of the space-time requires
$$16\Lambda^2-m^4=\left(\frac{4\pi}{G}\right)^2\frac{1}{n^2}$$
with $n$ an integer.

Finally, recall that this universe is compact, having topology $\mathbb S^1\times\mathbb S^3$, and the time for an observer at rest to traverse one cycle of the universe is given exactly by an elliptic integral that evaluates to
$$\frac{4\pi\sqrt{2}}{M(\sqrt{4\Lambda-m^2},\sqrt{4\Lambda+m^2})}$$
where $M$ is the Gauss \cite{Gauss} arithmetic-geometric mean.  In particular, for large $n$ at a fixed $m$, this is asymptotic to $\frac{8}{m}\log n$.

\section{Symmetry breaking}

Given this background metric, we can introduce the Robertson--Walker  scale factor as a function of the potential variable $\psi$,  exactly as in the standard Robertson--Walker theory.  So one possible scenario is that at the origin of the universe, at the highest possible temperatures we have a completely conformally invariant theory along the lines of work of Friedrich \cite{Friedrich1983} and Tod \cite{Tod2002}.  The physics here would be quite different from conventional physics.\footnote{See for example Penrose \cite{Penrose2012}, Witten \cite{Witten2004}, Sparling and Tillman \cite{Sparling2004}, and Sparling \cite{Sparling2007}.}  As the universe cools, there would be a phase transition to the sector under discussion in this work, dominated by the scalar field and accounting for the acceleration of the universe as discovered by Permutter {\em et al} \cite{Perlmutter1999} and Riess {\em et al} \cite{Riess1998}.  To match the present theory with physics, one needs to put in a matter model, and apply the renormalization group techniques originated by Wilson \cite{Wilson1971}, Becchi, Rouet, Stora, and Tyutin \cite{BRS1974}, \cite{BRS1975}, \cite{BRS1976}, \cite{Tyutin1975}, and Albrecht and Steinhardt \cite{Steinhardt1981}, \cite{AS1982}. Further cooling would allow a second phase transition allowing the development of the Robertson--Walker scale factor leading to the universe as we know it (see Weinberg \cite{Weinberg2008}, Chudnovsky, Frive, and Linde \cite{CKL1976}, Khlopov, Malomed, and Zeldovich \cite{KMZ1985}).

The scalar dominated phase also allows for a bridge of supersymmetric type connecting the conformal level of the presumably primordial Paneitz equation, with the conformal level of the more standard massless particles.\footnote{See Ramond \cite{Ramond1971}, Neveu and Schwarz \cite{Neveu1971}, Gervais and Sakita \cite{Gervais1971}, Golfand and Likhtman \cite{Golfand1972}, Volkov and Akulov \cite{Volkov1972}, \cite{Volkov1973}, \cite{Volkov1974}, and Wess and Zumino \cite{Wess1974}.}  The supersymmetric operator in question is the following:
\[ Q = \overline{Y}(Z, dZ), \hspace{10pt} Q^2 =0.\]
This operator naturally embeds by multiplication the sheaf cohomology describing massless particles (these are sheaf cohomology groups with values in the sheaf of germs of holomorphic \emph{functions} on twistor space) into the sheaf cohomology of one-forms on twistor space.  (This is a cohomological supersymmetry in the sense of Deligne and Morgan \cite{Deligne}.) The basic example maps the cohomology with values in the sheaf of germs of holomorphic functions homogeneous of degree minus two, which represents traditional scalar massless particles, by multiplication by the operator $Q$ to the sheaf cohomology of one-forms on projective twistor space, which encodes the Paneitz equation.  Conversely given the sheaf cohomology of one-forms, we restrict to the sheaf cohomology for functions by imposing the supersymmetry $Q = 0$.  So, in contrast to the prevailing view on the nature of supersymmetry, here the supersymmetry \emph{reduces} symmetry, indeed is responsible for the breaking of conformal invariance and setting a mass scale, allowing physics as we know it to exist.  

Finally note that we have \emph{two} such supersymmetric operators, the operator $Q$ and its adjoint $\overline{Q} = Y(\partial_Z, \delta_Z)$, where $\partial_Z$ is the dual-twistor-valued differential operator with respect to $Z$ and $\delta_Z$ is the dual-twistor-valued derivation of forms, of degree minus one, dual to multiplication by  $dZ$.  The full supersymmetry algebra is then:
\[ Q^2 =0,\hspace{10pt}  \overline{Q}^2 = 0, \]
\[ Q\overline{Q} + \overline{Q}Q = P_Y, \]
\[ [P_Y, Q] = 0, \hspace{10pt} [ P_Y, \overline{Q}] = 0, \]
\[ P_Y =  E_Y(\partial_Z,  Z)  -  E_Y(dZ,  \delta_Z).  \]
The operator $P_Y$ is the supersymmetric generalization of the operator $E_Y$.   So the theory is naturally supersymmetric.

From the viewpoint of mathematical physics, the present work has opened up the possibility of a purely gauge theoretic and geometric explanation for the origin of mass in conformally invariant gauge theories (Atiyah, Drinfeld, Hitchin, and Manin \cite{Atiyah1978}, Atiyah \cite{Atiyah1979}, Ward \cite{Ward1977}).

\end{document}